Ginzburg-Landau calculations of d-wave superconducting dot in s-wave superconducting matrix

Masaru Kato[a,b], Masayuki Ako[a,b], Masahiko Machida[b,c], Tomio Koyama[b,d], Takekazu Ishida[b,e]

[a] Department of Mathematical Sciences, Osaka Prefecture University, 1-1 Gakuencho, Sakai, Osaka 599-8531, Japan

[b] CREST, JST, 4-1-8, Honcho, Kawaguchi, Saitama 332-0012, Japan

[c] CCSE, Japan Atomic Energy Research Institute, 6-9-3 Higashi-Ueno, Taito-ku, Tokyo 110-0015 Japan

[d] Institute for Materials Research, Tohoku University, Sendai 980-8577, Japan

[e] Department of Physics and Electronics, Osaka Prefecture University, 1-1 Gakuencho, Sakai, Osaka 599-8531, Japan

Abstract

We have developed a numerical method that calculated superconducting states and magnetic field distributions for the composite structures of the High-$T_c$ superconductor and the conventional superconductor in arbitrary geometries. We show spontaneous magnetic flux appears at the corner of the boundary of these two superconductors. Also we propose High-Tc superconducting dot embedded in conventional superconductors, which is named as d-dot and show the spatial distribution of superconducting order parameters and the magnetic field.

*Corresponding Author

Dr. Masaru Kato

Postal Address: Department of Mathematical Sciences, Osaka Prefecture University, 1-1 Gakuencho, Sakai, Osaka 599-8531, Japan

Phone: +81-72-254-9368

Fax: +81-72-254-9916

E-mail address: kato@ms.osakafu-u.ac.jp

## 1. Introduction

Anisotropic superconductors show peculiar phenomena. High temperature superconductors are the most typical anisotropic superconductors. Its paring symmetry confirmed as $\hat{k}_x^2 - \hat{k}_y^2$ by the phase sensitive corner junction experiments [1-3]. In these experiments, Josephson weak links are made between an s-wave superconductor and two orthogonally oriented ac- and bc plane faces of the high-$T_c$ superconductor as shown in Fig. 1 (a). Because of the π phase shift between two weak links that comes from the sign change of the d-wave order parameter, a half-flux quantum vortex appears at the corner of the junction. This half-flux quantum vortex appears spontaneously under zero magnetic field. Also this state breaks time reversal symmetry, therefore this state is twice time degenerate. This means flux with opposite direction also can appear and these two states are equally stable.

This peculiar phenomenon leads to interesting problem that if the d-wave superconductor embedded in the conventional superconductor, then what happens? The size and shape of d-wave superconductor may affect the results. Especially, the shape of the high-Tc superconductor is square and the edges are parallel to the crystal axes (see Fig. 1(a)), a spontaneous magnetic flux might be expected. Recently, Hilgenkamp et al. showed that a zigzag boundary of the high-Tc superconductor and the s-wave conventional superconductor exhibit the antiferromagnetically ordered magnetic fluxes under zero magnetic field [4]. In contrast to this experiment, we consider a confined d-wave superconductor embedded in the s-wave superconducting matrix. In such configuration have potential use for some device. In this report we show a phenomenological calculation of this composite structure of the d-wave and s-wave superconductors. We use finite element method for solving phenomenological Ginzburg-Landau (GL) equation and investigate the superconducting state and magnetic field distribution in several geometries.

In section 2, we discuss our method, and in section 3 we show the numerical results. Section 3 is devoted to the discussion and summary.

**2. Method**

To investigate the field distribution in the composite structures of s- and d-wave superconductors, we start from the phenomenological GL theory. For the appearance of the spontaneous magnetic flux, the anisotropy of the d-wave superconductivity and interference of d- and s-wave superconductors are essential. Therefore we use the two-component GL free energy for d-wave superconductors, which is derived by Ren et al. from microscopic Gor'kov equation [5-7]. It contains s-wave component as well as d-wave component of the order parameter. From this free energy, four-fold symmetry of the vortex structure in High-Tc superconductors deduced [5-9]. Although including higher order terms, such four-fold symmetry of the vortex was shown for the purely d-wave superconductivity by Enomoto et al. [10] and Shiraishi et al. [11]. Similar free energy is used for spontaneous existence of the fractional magnetic flux at the grain boundary [12,13], and Josephson juctions between unconventional superconductors [14]. Adding a gauge fixing term $(1/8\pi)(\mathrm{div}\,A)^2$ and subtracting some constant terms, we use following free energy,

$$F_d(\psi_d,\psi_s,\boldsymbol{A}) = \int_\Omega d\Omega\, \alpha \lambda_d^5 \left\{ \frac{3}{8}\left[|\psi_d|^2 - |\psi_d^0(T)|^2\right]^2 + \left(\frac{\lambda_s}{\lambda_d}\right)^4 \left[|\psi_s|^2 + |\psi_s^0|^2\right]^2 \right.$$

$$+ \left(\frac{\lambda_s}{\lambda_d}\right)^2 \left[2|\psi_d|^2|\psi_s|^2 + \frac{1}{2}\left(\psi_s^{*2}\psi_d^2 + \psi_s^2\psi_d^{*2}\right)\right]$$

$$+ \frac{3}{4}\xi_d^2 \psi_d^0(T)^2 \left[|\Pi\psi_d^*|^2 + 2\left(\frac{\lambda_s}{\lambda_d}\right)^2 |\Pi\psi_s^*|^2 + \frac{\lambda_s}{\lambda_d}\left(\Pi_x^*\psi_s\Pi_x\psi_d^* - \Pi_y^*\psi_s\Pi_y\psi_d^* + \text{H.c.}\right)\right]$$

$$\left. + \kappa_d^2 \xi_d^2 \psi_d^0(T)^4 \left[\left(\text{rot}\,\tilde{\boldsymbol{A}} - \frac{2\pi}{\Phi_0}\boldsymbol{H}\right)^2 + (\text{div}\,\tilde{\boldsymbol{A}})^2\right]\right\}.$$

(1)

Here $\alpha = 7\zeta(3)/8(\pi T)^2$, $\lambda_d = V_d N(0)$ and $\lambda_s = V_s N(0)$ where $V_d$ ($V_s$) is an interaction constant for the d-(s-) wave paring, respectively, and $N(0)$ is the density of states at the Fermi energy. In this free energy, a repulsive interaction for the s-wave Cooper pairing is assumed because strong Coulomb interaction prevent the s-wave superconductivity in high-Tc superconductors. $\psi_d = \Delta_d/\lambda_d$ and $\psi_s = \Delta_s/\lambda_s$ are d- and s-wave components of the superconducting order parameter and $\psi_d^0(T)$ is a uniform d-wave order parameters at temperature $T$, and $\psi_s^0$ is a constant that is determined from the repulsive interaction. Also $\Pi = \left(\frac{1}{i}\nabla - \tilde{\boldsymbol{A}}\right)$ and $\tilde{\boldsymbol{A}} = \frac{2\pi}{\Phi_0}\boldsymbol{A}$ is the normalized vector potential, where $\Phi_0 = \frac{hc}{2e}$ is a flux quantum. $\xi_x$ and $\kappa_d$ are the GL coherence length and the GL-parameter of the d-wave superconductors, respectively. The anisotropy comes from the gradient term $\left(\Pi_x^*\psi_s\Pi_x\psi_d^* - \Pi_y^*\psi_s\Pi_y\psi_d^* + \text{H.c.}\right)$, which couples spatial variations of s-wave and d-wave order parameters. Therefore anisotropy appears with non-uniform superconducting state.

For our composite structure, we must take into account the interference between the s-wave and d-wave order parameters also in the s-wave superconductors. Therefore we also consider the d-wave component of the order parameter in the s-wave superconductors. We assume that the interaction for the d-wave pairing in the s-wave superconductor is repulsive. Then the GL free energy for the s-wave superconductor is given as,

$$F_s(\psi_d,\psi_s,\boldsymbol{A}) = \int_\Omega d\Omega \frac{\alpha \lambda_s^5}{2} \Bigg\{ \Big[|\psi_s|^2 - |\psi_s^0(T)|^2\Big]^2 + \frac{3}{8}\left(\frac{\lambda_d}{\lambda_s}\right)^4 \Big[|\psi_d|^2 + |\psi_d^0|^2\Big]^2$$

$$+ \left(\frac{\lambda_s}{\lambda_d}\right)^2 \Big[2|\psi_d|^2|\psi_s|^2 + \frac{1}{2}\left(\psi_s^{*2}\psi_d^2 + \psi_s^2\psi_d^{*2}\right)\Big]$$

$$+ \xi_s^2 \psi_s^0(T)^2 \Bigg[\left(\frac{\lambda_d}{\lambda_s}\right)^2 |\boldsymbol{\Pi}\psi_d^*|^2 + 2|\boldsymbol{\Pi}\psi_s^*|^2 + \frac{\lambda_d}{\lambda_s}\left(\Pi_x^*\psi_s \Pi_x \psi_d^* - \Pi_y^*\psi_s \Pi_y \psi_d^* + \text{H.c.}\right)\Bigg]$$

$$+ 2\kappa_s^2 \xi_s^2 \psi_s^0(T)^4 \Bigg[\left(\text{rot}\,\tilde{\boldsymbol{A}} - \frac{2\pi}{\Phi_0}\boldsymbol{H}\right)^2 + \left(\text{div}\,\tilde{\boldsymbol{A}}\right)^2\Bigg]\Bigg\}.$$

(2)

Here, meanings of symbols are same as in Eq. (1). $\xi_x$ and $\kappa_d$ are GL coherence length and the GL-parameter in the s-wave superconductor. $\psi_s^0(T)$ is a uniform d-wave order parameters at temperature $T$, and $\psi_d^0$ is a constant that is determined from the repulsive interaction. Added to these free energies, at the boundary between d- and s-wave superconductors, we impose the continuity conditions of both of the order parameters.

In order to minimize these free energies Eqs. (1) and (2), we use the finite element method (FEM) [15]. The FEM was applied for solving the GL equation for s-wave superconductors in periodic vortex structures [16]. Also vortex structures in d-wave superconductors were investigated by FEM [17,18]. Because this method is applicable to arbitrary finite geometries, so we can design the composite structures and calculated superconducting state at will.

In the following, we only consider two-dimensional configurations, therefore we divide all regions of the superconductor into triangular elements. Then in the each element, we expand order parameters and the vector potential by so called area coordinates as,

$$\psi_s(\mathbf{r}) = \sum_{ie} \psi_{si}^e N_i^e(\mathbf{r}), \tag{3}$$

$$\psi_d(\mathbf{r}) = \sum_{ie} \psi_{di}^e N_i^e(\mathbf{r}), \tag{4}$$

$$\mathbf{A}(\mathbf{r}) = \sum_{ie} \mathbf{A}_i^e N_i^e(\mathbf{r}), \tag{5}$$

where $N_i^e(\mathbf{r})$ is fthe i-th area coordinate in the e-th element, which is given as,

$$N_i^e(x,y) = \frac{1}{2S_e}(a_i + b_i x + c_i y), \tag{6}$$

where $S_e$ is the area of the e-th element and $a_i$, $b_i$ and $c_i$ are constants that are determined from the coordinates of three vertices of the triangular element. $\psi_{si}^e$, $\psi_{di}^e$ and $A_{di}^e$ are values at the i-th vertex in the e-th element. Substituting these expansions to the free energy and minimizing it we obtain coupled non-linear equations for $\psi_{si}^e$, $\psi_{di}^e$ and $A_{di}^e$. Linearizing non-linear term and using iteration method, we solved them self-consistently.

## 3. Results

First we consider corner junction geometry. In Fig.2 we show the order parameter distributions and the magnetic field distribution in zero external magnetic field. We take $\kappa_d = 3.0$, $\kappa_s = 2.0$, $\xi_d = 0.1$, $\xi_s = 0.1$ and $T_{cd}/T_{cs} = 2.0$, where $T_{cd}$ and $T_{cs}$ are critical temperature for d- and s- wave superconductors, respectively. Temperature is set as $T/T_d = 0.1$. In agreement with the experiment, there exists spontaneous magnetic field at the right-angled corner. The d-wave order parameter penetrates into the s-wave SC region slightly, and vice versa. But it is sufficient for inducing the magnetic flux.

In Fig. 3, we show the results for the zigzag boundary between d- and s-wave superconductors. The magnetic fluxes appear at two right angled corners and directions of the fluxes are opposite. This is because there is a repulsive interaction between vortices in same direction and an attractive interaction between them in opposite direction. This result also agrees with the experiment of Hilgenkamp et al. [3].

Next we consider a square shaped d-wave superconductor embedded in the s-wave superconductor (square d-dot). In Figs. 4 and 5, we show distributions of order parameters and the magnetic field for different size of d-dot. For both cases, magnetic fluxes appear at the every corner and they are ordered antiferromagnetically. The s-wave order parameter penetrates into the d-wave SC region, compare to the corner junction or the zigzag geometries, especially for smaller d-dot. This comes from the finiteness of the d-wave SC reigon. For smaller size (Fig.5), magnetic field nearly mixes, but for the larger d-dot (Fig.4) magnetic field is well separated. This tendency implies that for the bigger d-dot, the four magnetic flux appears independently, and for smaller d-wave superconductors, magnetic field may disappear.

## 4. Summary

We have developed the numerical method for investigate the superconducting state and the field distribution in the composite structures of d- and s-wave superconductors. Our results reproduce experiments on the corner junction [1-3] and the zigzag boundary [4]. Moreover, we have investigate a finite d-wave superconductor in the s-wave superconductor, which we call as d-dot. Here we have considered only square shaped d-dots. We have also investigated the d-dots in the other shapes [19].

**Acknowledgements**

**References**

[1] Wollman et al., Phys. Rev. Lett. 71 (1993) 2134.

[2] D. J. Van Harlingen, Rev. Mod. Phys. 67 (1995) 515.

[3] C. C. Tsuei and J. R. Kirtley, Rev. Mod. Phys. 72 (2000) 969.

[4] H. Hilgenkamp et al., Nature **422** (2003) 50.

[5] Y. Ren, J-H. Xu, and C. S. Ting, Phys. Rev. Lett. **74** (1995) 3680.

[6] J.H. Xu, Y. Ren, and C. S. Ting, Phys. Rev. B 52 (1995) 7663.

[7] J. H. Xu, Y. Ren, and C. S. Ting, Phys. Rev. B 53 (1996) 2991.

[8] A. J. Berlinsky et al., Phys. Rev. Lett. 75 (1995) 2200.

[9] M. Franz, et al., Phys. Rev. B 53 (1996) 5795.

[10] N. Enomoto, M. Ichioka, and K. Machida, J. Phys. Soc. Jpn. 66 (1997) 204.

[11] J. Shiraishi, M. Kohmoto, and K. Maki, Phys. Rev. B 59 (1999) 4497.

[12] M. Sigrist, D. B. Bailey, and R. B. Laughlin, Phys. Rev. Lett. 74 (1995) 3294.

[13] D. B. Bailey, M. Sigrist, and R. B. Laughlin, Phys. Rev. B 55 (1997) 15239.

[14] K. Kuboki and M. Sigrist, J. Phys. Soc. Jpn. 65 (1996) 361.

[15] Q. Du, M. D. Gunzburger, and J. S. Peterson, SIAM Rev. **34** (1992) 54.

[16] Q. Du, M. D. Gunzburger, and J. S. Peterson, Phys. Rev. B **46** (1992) 9027.

[17] Z. D. Wang and Q. H. Wang, Phys. Rev. B **55** (1997) 11756.

[18] Q. Li, Z. D. Wang, Q. H. Wang, Phys. Rev. B **59** (1999) 613.

[19] M. Ako et al., in this issue.

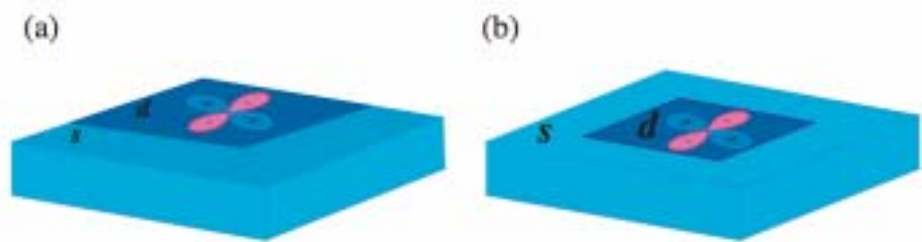

Fig.1

Schematic diagram of the composite structure of d-wave and s-wave superconductors. (a) Corner junction geometry. (b) Square shaped d-wave superconductor embedded in the s-wave matrix. All of the sides of the d-wave superconductor are parallel to a or b axis.

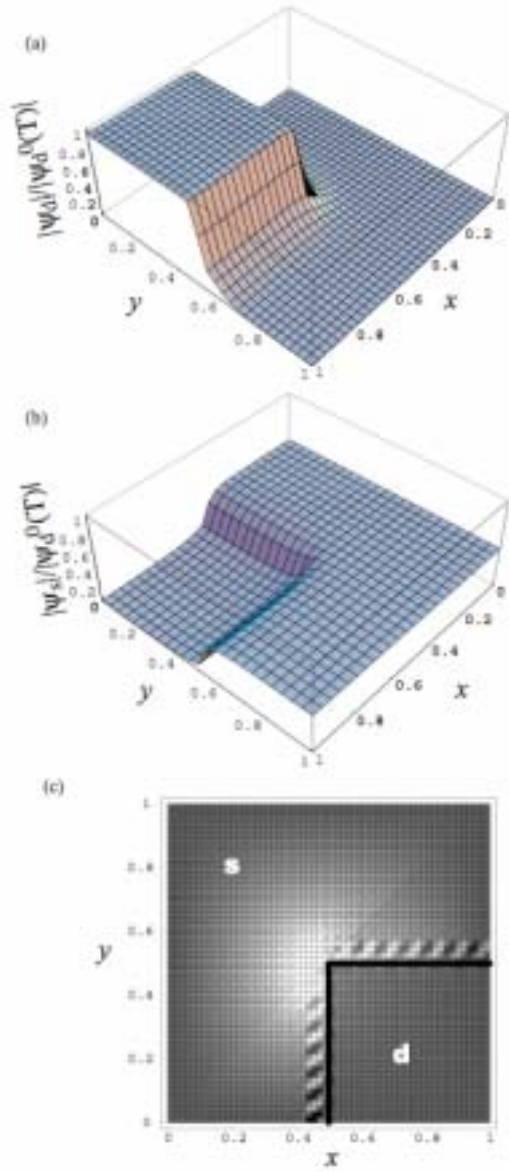

Fig. 2

Spatial distributions of the amplitude of the s-wave (a), d-wave (b) order parameter and magnetic field (c) for the corner junction geometry. The strength of the magnetic field is represented in the gray scale, where white means the magnetic field is largest and upward and black means the magnitude of the magnetic field is largest but direction is downward.

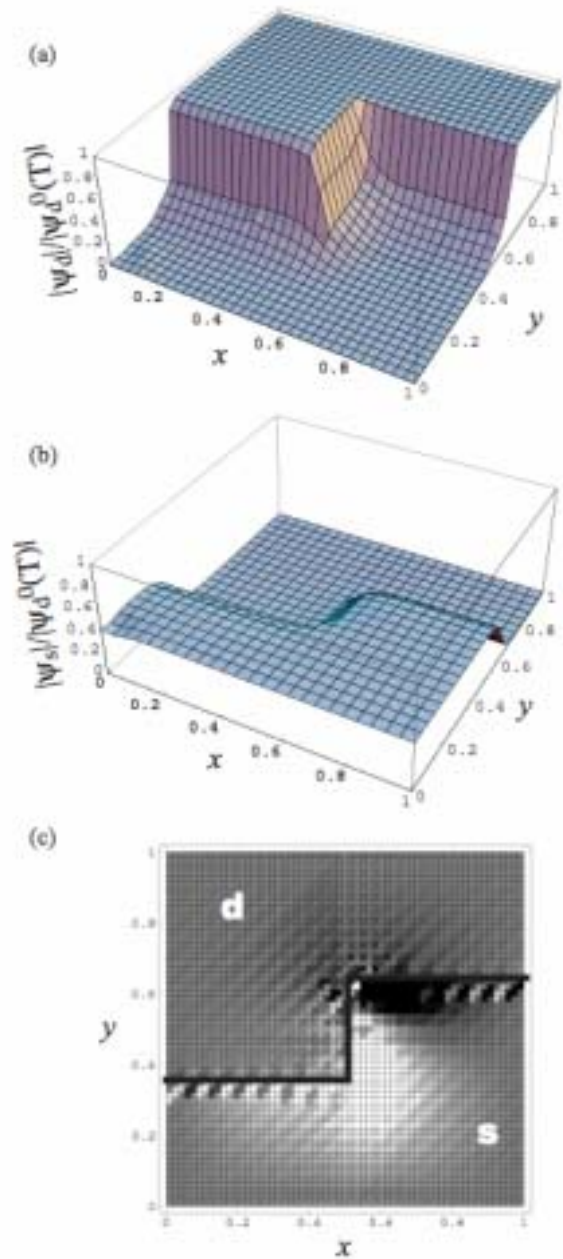

Fig.3

Spatial distributions of the amplitude of the s-wave (a), d-wave (b) order parameter and magnetic field (c) for the staircase boundary. Meanings of the symbols are same as that of Fig.2.

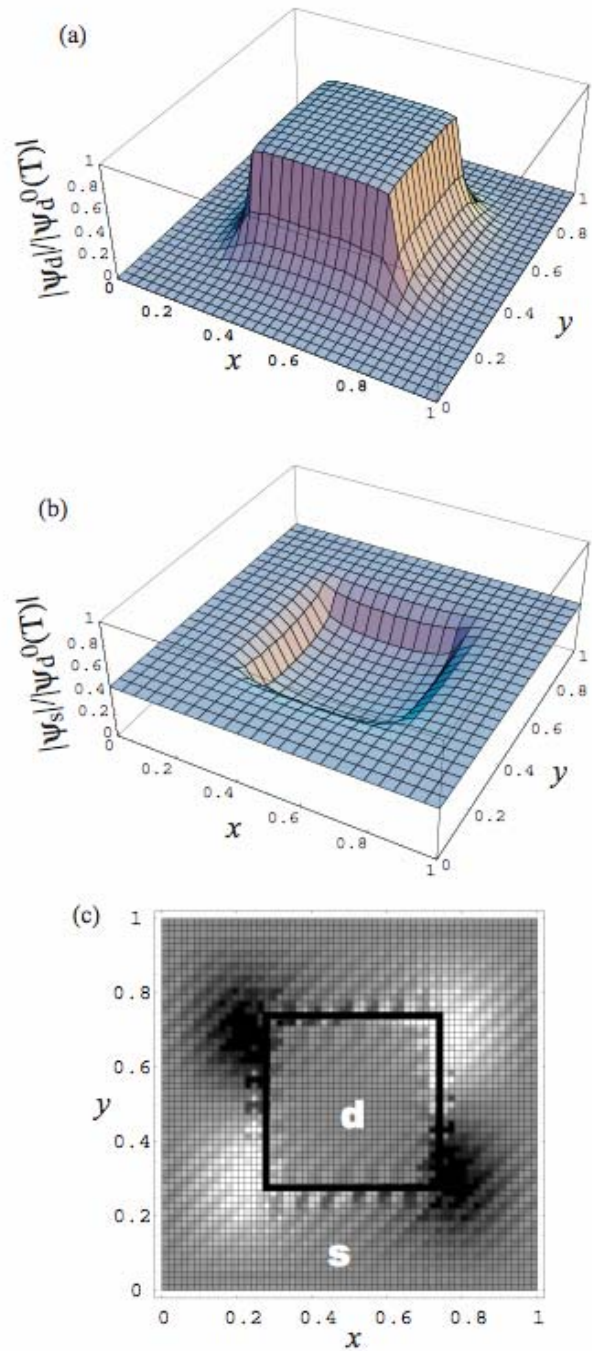

Spatial distributions of the amplitude of the s-wave (a), d-wave (b) order parameter and magnetic field (c) for the d-dot. Meanings of the symbols are same as that of Fig.2.

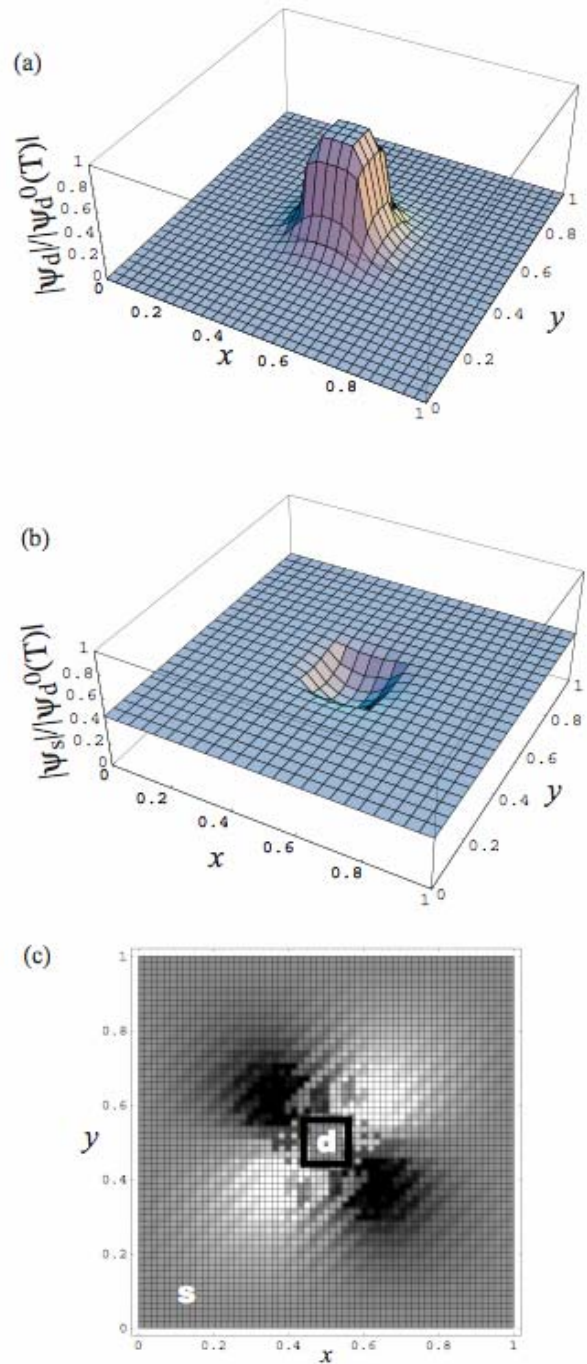

Fig.5

Spatial distributions of the amplitude of the s-wave (a) and d-wave (b) order parameters and magnetic field (c) for the d-dot. Meanings of the symbols are same as that of Fig.2.